# Symmetry-enforced three-dimensional Dirac phononic crystals

Xiangxi Cai,[1†] Liping Ye,[1*] Chunyin Qiu,[1] Meng Xiao,[1] Rui Yu,[1] Manzhu Ke,[1] Zhengyou Liu[1,2*]

[1]Key Laboratory of Artificial Micro- and Nano-Structures of Ministry of Education and School of Physics and Technology, Wuhan University, Wuhan 430072, China

[2]Institute for Advanced Studies, Wuhan University, Wuhan 430072, China

Dirac semimetals, the materials featured with discrete linearly crossing points (called Dirac points) between four bands, are critical states of topologically distinct phases. Such gapless topological states have been accomplished by a band-inversion mechanism, in which the Dirac points can be annihilated pairwise by perturbations without changing the symmetry of the system. Here, we report an experimental observation of Dirac points that are enforced completely by the crystal symmetry, using a nonsymmorphic three-dimensional phononic crystal. Intriguingly, our Dirac phononic crystal hosts four spiral topological surface states, in which the surface states of opposite helicities intersect gaplessly along certain momentum lines, as confirmed by our further surface measurements. The novel Dirac system may release new opportunities for studying the elusive (pseudo)relativistic physics, and also offer a unique prototype platform for acoustic applications.

[*]Corresponding author. Email: lpye@whu.edu.cn; zyliu@whu.edu.cn (Z.L.)

Discovery of new topological states of matter has become a vital goal in fundamental physics and material science[1,2]. Three-dimensional (3D) Dirac semimetal (DSM)[3-13], accommodating many exotic transport properties such as anomalous magnetoresistance and ultrahigh mobility[14,15], is an exceptional platform for exploring topological phase transitions and other novel topological quantum states. It is also of fundamental interest to serve as a solid-state realization of (3+1)-dimensional Dirac vacuum. A DSM phase may appear accidentally at the quantum transition between normal and topological insulators[16,17]. The approach to such a single critical point demands fine-tuning of the alloy chemical compositions, which limits the experimental accessibility to the fascinating physics of 3D Dirac fermions. 3D DSMs can also emerge robustly over a range of Hamiltonian control parameters, as distinguished into two classes[3,4]. The first one, already realized in $Na_3Bi$[7,8] and $Cd_3As_2$[9,10], occurs due to band inversion[5,6]. The Dirac points, lying on generic momenta of a specific rotation symmetry axis, always come in pair and can be eliminated by their merger and pairwise annihilation through continuously tuning parameters[3,4] that preserve the symmetry of the system. The second class is featured with Dirac points pinned stably to discrete high-symmetry points on the surface of the Brillouin zone (BZ). Markedly different from the first class of DSMs, the occurrence of Dirac points is an unavoidable result of the nonsymmorphic space group of the material[11-13], which cannot be removed without changing crystal symmetry. Although some solid-state candidate materials have been proposed[4,11,12], the symmetry-enforced 3D DSMs have never been experimentally realized because of the great challenge in synthesizing materials[4,7].

Recently, numerous distinct topological states have been demonstrated in classical wave systems[18,19] such as photonic crystals[20-28] and phononic crystals[29-34], which offer opportunities for exploring topological physics in a fully controllable manner. Here we report the first experimental realization of a 3D phononic crystal that hosts symmetry-enforced Dirac points at the BZ corners. The fourfold degeneracy is protected by a nonsymmorphic space group that couples point operations (rotations and mirrors) with nonprimitive lattice translations. In addition to the Dirac points identified directly by angle-resolved transmission measurements, highly intricate quad-helicoid surface states have been unveiled by our surface measurements and associated Fourier spectra. Specifically, the surface states are composed of four gaplessly-crossed spiral branches[13], and thus are strikingly different from the double Fermi arc surface states observed recently in electronic[8] and photonic systems[28]. Excellent agreements are found between our experiments and simulations.

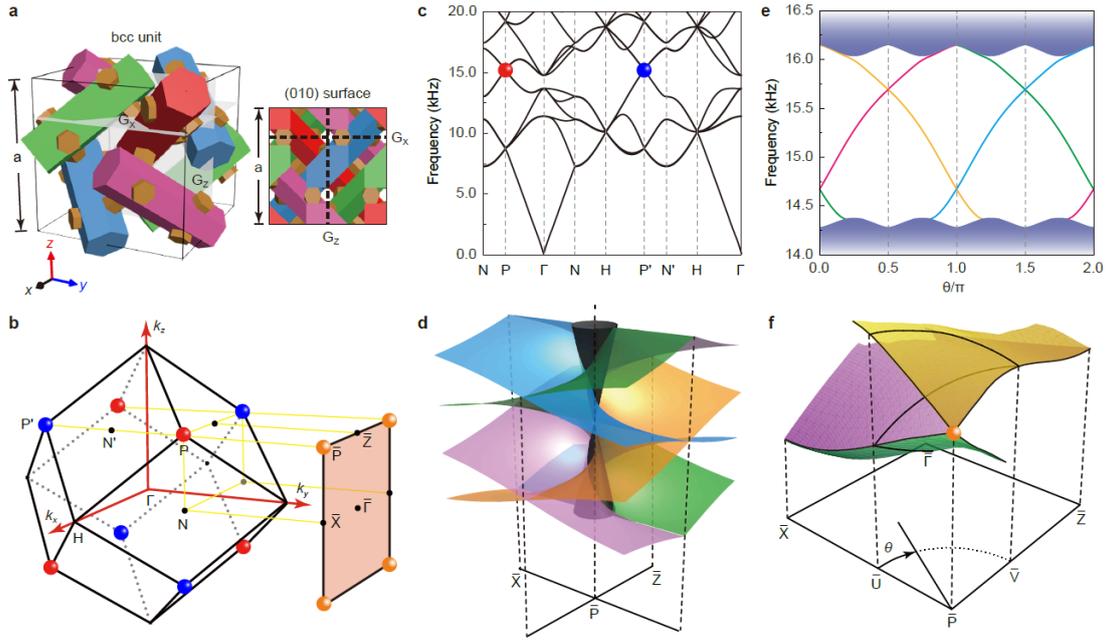

**Figure 1 | Symmetry-enforced Dirac points and quad-helicoid topological surface states in a nonsymmorphic phononic crystal. a**, Schematics of the bcc unit (left panel) of the phononic crystal and its (010) surface (right panel) featured with two glide mirrors $G_x$ and $G_z$. **b**, 3D bcc BZ and its (010) surface BZ. The colored spheres highlight the bulk Dirac points with equal frequency and their projections onto the surface BZ. **c**, Bulk bands simulated along several high-symmetry directions. **d**, Schematic of the quad-helicoid surface state dispersions (color surfaces), where the grey cone labels the projection of bulk states. **e**, Surface bands simulated along a circular momentum loop of radius $0.4\pi/a$ (as shown in **f**) centered at $\bar{P}$. The shadow regions indicate the projected bulk states. (**f**) 3D plot of the surface dispersion simulated in the first quadrant of the surface BZ. Bulk band projections are not shown for clarity.

As illustrated in Fig. 1a, our Dirac phononic crystal has a body-centered-cubic (bcc) lattice, associated with lattice constant $a = 2.8\,\text{cm}$. The main body of the building block consists of four inequivalent resin cylinders, which are labeled with different colours and oriented along different bcc lattice vector directions. All the cylinders have a regular hexagonal cross-section of sidelength 0.42 cm. To facilitate the sample fabrication, these cylinders are connected with short hexagonal bars of sidelength 0.21 cm. The rest of the volume is filled with air. Numerically, the photosensitive resin material used for printing the acoustic structure is treated as rigid and sound propagates only in air (at speed 342m/s), considering the great acoustic impedance mismatch between resin and air.

The crosslinked network structure belongs to the nonsymmorphic space group 230 ($Ia\bar{3}d$), featured with inversion symmetry and multiple screw rotations and glide reflections. The crystal symmetry enables rich point and line degeneracies (see Supplementary Materials). Interestingly, the little group at P and P′, a pair of time-reversal related Brillouin zone (BZ) corners (Fig. 1b), has 24 group elements and supports only fourfold degeneracy. This is confirmed by the band structure in Fig. 1c, which is stabilized with two distinct kinds of Dirac points at P (P′). The first kind of Dirac points, crossed with bands of different slopes and thus called generalized Dirac points[22] (e.g. the lowest ones at P and P′ in Fig. 1c), corresponds to a four-dimensional irreducible representation, whereas the second kind, crossed with bands of identical slopes, corresponds to two inequivalent two-dimensional irreducible representations stuck with time-reversal symmetry. Hereafter we focus on the latter case (as specified with color spheres in Fig. 1c), around which the bands are rather clean and carry a wide frequency window of linear dispersion. The system can be captured by a simple four-band effective Hamiltonian derived from $k \cdot p$ theory,

$$\mathcal{H} = \begin{pmatrix} O & H \\ H^\dagger & O \end{pmatrix}, \text{ where } H = \eta\left(\delta k_y \sigma_x - \delta k_x \sigma_y + \delta k_z \sigma_z\right), \eta \text{ is a complex parameter}$$

determined by the acoustic structure, $(\delta k_x, \delta k_y, \delta k_z)$ characterizes the momentum deviation from P, and $\sigma_i$ are Pauli matrices (see Supplementary Materials). The Dirac model gives isotropic linear dispersions around the Dirac point, which are much different from those anisotropic ones observed previoulsy[7-10,28]. A nontrivial $Z_2$ topological invariant, defined on a momentum sphere enclosing the Dirac points, can be used to depict the topology of such fourfold band clossing points[13]. It is derived by considering the pseudo anti-unitary symmetry ($\vartheta$) composited by a glide reflection ($G$) and time-reversal symmetry ($T$), i.e., $\vartheta = G*T$ with $\vartheta^2 = -1$. In addition, markedly different from the Dirac points created by band inversion[5,6], which can be annihilated pairwise without changing the crystal symmetry, here the Dirac points are guaranteed completely by the nonsymmorphic symmetries. The topological robustness of the Dirac points against symmetry-preserving perturbations has been identified numerically by two detailed examples (Supplementary Materials, Fig. S1).

Unlike Weyl semimetals that host topologically nontrivial Fermi arcs on their surfaces[35], the presence of topological surface states in a DSM is more subtle because the Dirac points carry a zero Chern number[3,4,36]. However, for a non-symmorphic DSM that has Dirac points featured with nontrivial $Z_2$ index, the band crossing points will pairwise connected by symmetry-protected Fermi arcs on the surface, associated with a unique connectivity determined by the $Z_2$ topological charge[13]. The dispersion of the topological surface states can be mapped to an intersecting multi-helicoid structure, where the intersections between the helicoids are protected from being gapped by the glide symmetries preserved on the specific surface. In our case, the Dirac phononic crystal supports elusive quad-helicoid surface states[13] if truncated with (010) surface or its equivalents, which can be characterized by the wallpaper group $p2gg$. Below we focus on the (010) surface that preserves the two glide mirrors $G_x = \{M_z | (a/2)\hat{x} + (a/2)\hat{z}\}$ and $G_z = \{M_x | (a/2)\hat{z}\}$ of the bulk crystal (Fig. 1a). For this specific crystal surface, the two inequivalent Dirac points are projected onto the four *equivalent* surface BZ corners $\bar{P}$ (Fig. 1b). As schematically illustrated in Fig. 1d, the quad-helicoid surface states are featured with two crucial signatures. First, there are totally four branches of spiral surface states for any given momentum loop enclosing $\bar{P}$: two with positive helicities and two with negative helicities. This can be seen in Fig. 1e, the gapless surface bands simulated along a circular loop centered at $\bar{P}$. Second, the surface states of opposite helicities intersect along certain momentum lines, in which the intersecting double degeneracies are protected by the glides $G_x$ and $G_z$ assisted with time-reversal symmetry[13]. This is exhibited clearly in the simulated global dispersion profile (Fig. 1f), which shows nodal line degeneracies along the surface BZ boundaries $\bar{P}\bar{X}$ and $\bar{P}\bar{Z}$. (Only 1/4 surface BZ is provided due to the presence of the two glides.) For a generic selection of the crystal surface, the nodal line degeneracy of the surface dispersion disappears due to the absence of the glide symmetries (Supplementary Materials, Fig. S2).

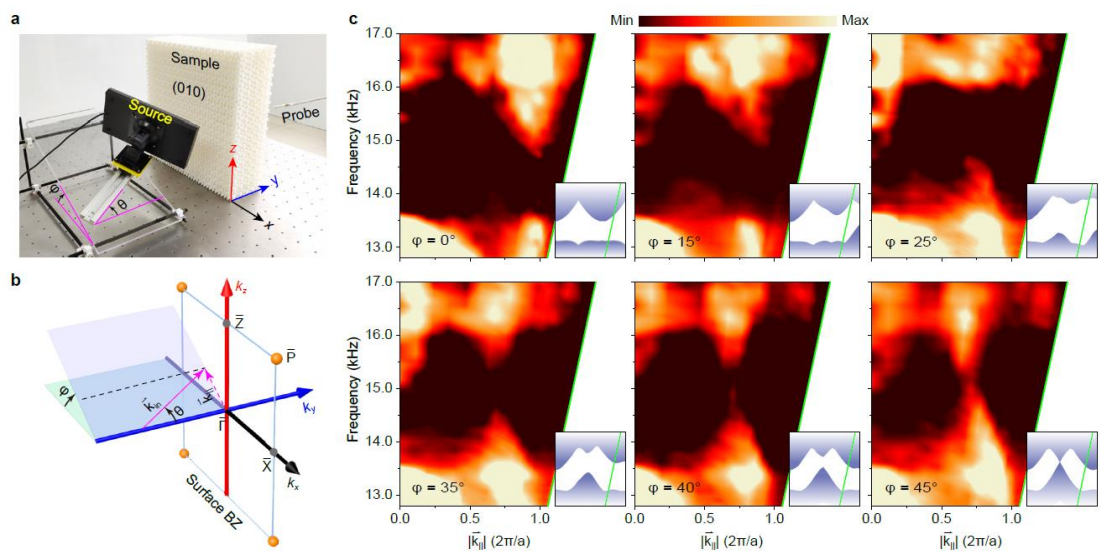

**Figure 2 | Experimental identification of the symmetry-enforced Dirac points. a**, Experimental setup for measuring sound transmission. **b**, Schematic of exciting bulk states according to the momentum conservation $\vec{k}_\parallel = \vec{k}_{\text{in}} \sin\theta$. **c**, $\theta$-resolved transmission spectra measured for different $\varphi$ values. The slanted boundary (green line) in each panel corresponds to the 'sound cone' $|\vec{k}_\parallel| = |\vec{k}_{\text{in}}|$, beyond which no transmission can be measured. Insets: Simulated bulk states (shadow regions) projected along the *y* direction, scaled to the same range and ratio as the measured

data.

The presence of symmetry-enforced Dirac points was confirmed by angle-resolved transmission measurements. Figure 2a demonstrates our experimental setup. The sample, fabricated precisely by 3D printing technique, has a size of 47.6 cm, 14.0 cm and 47.6 cm along the *x*, *y* and *z* directions, respectively. A rectangular acoustic horn was used to launch a collimated beam upon the (010) surface of the sample, where the incident direction can be characterized by the angles $\theta$ and $\varphi$. As illustrated in Fig. 2b, a bulk state is expected to be excited when its in-plane momentum $\vec{k}_{\parallel}$ matches that of the incident wavevector $\vec{k}_{in}\sin\theta$ at the same frequency. The transmitted sound signal was scanned by a 1/4 inch microphone (B&K Type 4958-A) and recorded by a multi-analyzer system (B&K Type 3560B). The averaged sound intensities were normalized to those measured in the absence of sample. The bulk states were mapped out by varying $\theta$ and $\varphi$. Here only $\varphi \in [0, 45°]$ was focused thanks to the multiple glide mirrors of the system. (For completeness, similar data for $\varphi \in [45°, 90°]$ are provided in Supplementary Materials, Fig. S3.) Specifically, at $\varphi = 45°$ the incident beam scans through the Dirac point. Figure 2c shows the transmission data measured for six representative $\varphi$ values, compared with the numerical bulk dispersions projected along the corresponding directions (insets). All the transmission spectra agree reasonably well with the numerical band structures, where the low transmission near the sound cone can be attributed to the smaller effective cross section area of the sample at large $\theta$. In particular, as expected in the case of $\varphi = 45°$, a conic touch is observed around 15.3 kHz in frequency and $0.71 \pi / a$ in wavevector. The point crossing is lifted gradually as $\varphi$ decreases from $45°$.

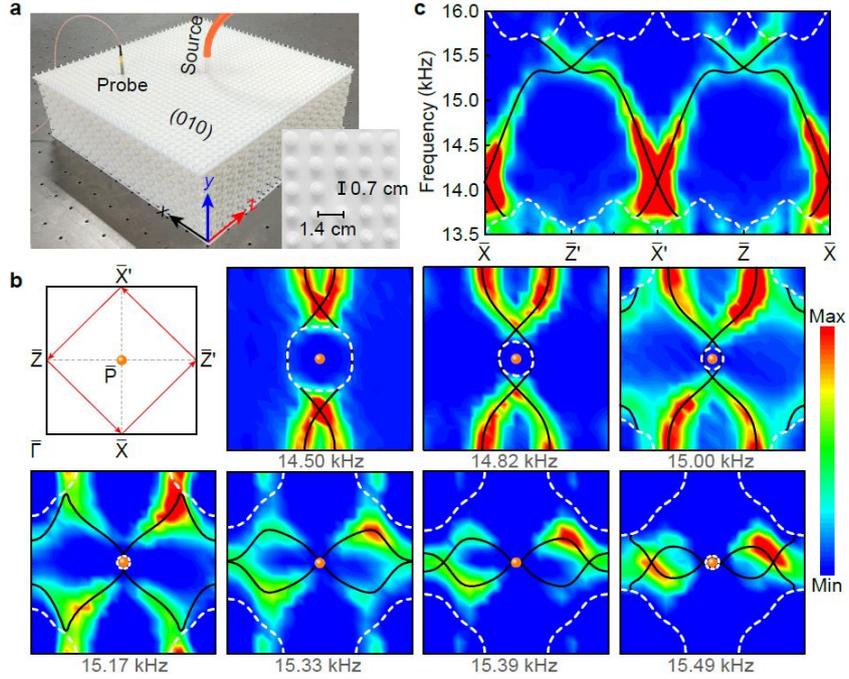

**Figure 3 | Experimental observation of quad-helicoid topological surface states. a**, Experimental setup for surface field measurements. The inset shows the details of the cover plate with circular holes opened or sealed. The plugs that seal the holes were opened one-by-one during the measurement. **b**, Isofrequency contours plotted in one surface BZ centered at $\bar{P}$ (see the first panel). The color scale shows the experimental data, comparing with the corresponding simulation results (black curves). The orange spheres label the projected Dirac points and the white dashed lines enclose the bulk band projections. **c**, Frequency dependent surface spectra (color scale) measured along the momentum path specified in the first panel of **b**.

Furthermore, we performed surface measurements to identify the highly intricate topological quad-helicoid surface states, which have not been experimentally observed in any topological system so far. Figure 3a shows our experimental setup. To mimic the rigid boundary condition involved in our simulations, an additional resin plate of thickness 0.2 cm was integrated on the (010) surface, which served as a trivial acoustic insulator to guarantee the presence of topological surface states. Since the typical air channels of the sample are too narrow to accommodate the sound source and probe directly, the plate was perforated with a square lattice of holes (see inset), where the lattice spacing 1.4 cm gives the in-plane scanning step. During our measurement, all the holes were sealed except those reserved for the sound source and detector. To excite surface states, a broadband point-like sound source, launched from a subwavelength-sized tube, was injected into one hole near the center of the sample surface. The localized surface field was detected hole-by-hole through a portable microphone. By Fourier transforming the surface pressure field, we mapped out the nontrivial surface arc for any desired frequency[31]. Figure 3b shows such data for a sequence of frequencies. As predicted by simulations, the measured surface arcs (bright color) exhibit clear crossings at the surface BZ boundaries $\bar{X}\bar{X}'$ and $\bar{Z}\bar{Z}'$. Our experimental results capture well the simulated isofrequency contours of the topological surface states (black lines), despite the band broadening due to the finite-size effect. Note that the amplitude signals of the bulk states (enclosing by white dashed lines) are much weaker than those of topological surface states which are highly confined to the surface. To further identify the gapless quad-helicoid surface states, we present the surface spectra (Fig. 3c) measured along the momentum loop specified in the first panel of Fig. 3b. Comparing with the loop used in Fig. 1e, this square loop enclosing $\bar{P}$ point is bigger and favored to demonstrate the gapless intersection of the surface bands in a wide bulk gap. As expected, two pairs of surface bands with opposite helicities traverse the bulk gap and cross stably at the high-symmetry momenta $\bar{X}$ ($\bar{X}'$) and $\bar{Z}$ ($\bar{Z}'$). Again, an excellent agreement is found between our experiment and simulation.

In conclusion, we have constructed and identified a spinless Dirac crystal working for airborne sound, which exhibits highly intricate properties in both the bulk and surface states, in sharp contrast to those realized previously in condensed matter systems[7-10]. Note that a very recent work by S. Zhang *et al*. has made the first step toward the experimental study of 3D Dirac points in classical wave systems[28]. Interestingly, the Dirac points are constructed by electromagnetic duality symmetry (which are unique in electromagnetic systems), also strikingly different from the crystalline symmetry involved here. Starting with our structure, one can design various interesting 3D acoustic topological states (e.g., Weyl points[29,31] and line nodes[37,38]) through symmetry reduction. Here we would like to mention that, more or less, there is dissipation loss in our airborne phononic crystal, which may introduce a non-Hermitian modification in the effective Hamiltonian. The influence of the non-Hermiticity on our system deserves to be explored further in the future. This study may open up new manners for controlling sound, such as realizing unusual sound scattering and radiation, considering the conical dispersion and vanishing density of states around the Dirac points. Last but not least, the dispersion around the Dirac point is isotropic and thus our macroscopic system serves as a good platform to simulate relativistic Dirac physics.

**Methods**

**Numerical simulations.**

All simulations were performed using COMSOL Multiphysics, a commercial solver package based on the finite-element method. The bulk band structure in Fig. 1c was calculated by a single unit cell imposing with specific Bloch boundary conditions. Similar calculations gave the projected bulk states along the $y$ direction (Fig. 2c, shadowed region). A ribbon structure was used to calculate the surface band for a desired surface (Figs. 1e-1f and Figs. 3b-3c), imposing with Bloch boundary conditions along the $x$ and $z$ directions, and a rigid boundary condition along the $y$ direction, respectively. The ribbon was long enough to avoid the coupling between the opposite surfaces. Surface states were distinguished from the projected bulk states by inspecting the surface localizations of eigenstates.

**Experimental measurements.**

Our experiments were performed for airborne sound at audible frequency. The slab-like sample, consisting of $17 \times 5 \times 17$ structural units along the $x$, $y$ and $z$ directions, was prepared by photosensitive resin via 3D printing. The macroscopic characters of our acoustic system enable precise sample fabrication and less demanding signal detection. To excite the bulk states, a rectangular acoustic horn (of surface area 24.0 cm × 10.0 cm) was used to launch Gaussian beams at controllable orientations (Fig. 2a), whereas a narrow tube (of diameter 0.8 cm) was used to export point-like sound signal to excite the topological surface states (Fig. 3a). During both measurements, a portable microphone was moved on the $x$-$z$ plane to scan the pressure fields, together with another identical microphone fixed for phase reference. Both the amplitude and phase information of the input and output signals, swept from 11.8 kHz to 18.2 kHz with an increment of 0.032 kHz, were recorded and analyzed by a multi-analyzer system. To map out each surface arc of given frequency (Fig. 3b), two-dimensional Fourier transformation was performed to the scanned surface field; this further gave the frequency dependent surface spectra along the specific momentum loop (Fig. 3c).

**Data availability**

The data that support the plots within this paper and other findings of this study are available from the corresponding author upon reasonable request.

**Code availability**

All codes that support this study are available from the corresponding author upon reasonable request.

**Acknowledgements**

We thank Fan Zhang for fruitful discussions. This work is supported by the National Basic Research Program of China (Grant No. 2015CB755500); National Natural Science Foundation of China (Grant Nos., 11774275, 11890701, 11674250 and 11534013); National Key R&D Program of China (Grant 2018YFA0305800); Natural Science Foundation of Hubei Province (Grant No. 2017CFA042); and the Young Top-notch Talent for Ten Thousand Talent Program (2019- 2022).

**Author contributions**

Z.L. conceived the original idea and supervised the project. X.C. designed the acoustic structure and performed the simulations. L.Y. fabricated the sample and made the measurements. C.Q., X.C., L.Y., M.X., R.Y., M.K. and Z.L. analyzed the data and wrote the manuscript. All authors contributed to scientific discussions of the manuscript.

**Additional information**

Supplementary information is available in the online version of the paper. Reprints and permission information are available online at www.nature.com/reprints. Correspondence and requests for materials should be addressed to L.Y. and Z. L.

**Competing Interests**

The authors declare that they have no competing financial interests.